\documentclass[12pt]{iopart}
\expandafter\let\csname equation*\endcsname\relax
\expandafter\let\csname endequation*\endcsname\relax
\usepackage{amsmath}
\usepackage{amsfonts}
\usepackage{amssymb}
\usepackage{mathrsfs}
\usepackage{graphicx}
\usepackage{color}
\usepackage{float}
\usepackage{mathtools}
\usepackage{algorithm}
\usepackage{algpseudocode}
\usepackage{wrapfig}
\usepackage{mathbbol}
\usepackage[colorlinks, linkcolor=blue, citecolor = blue, urlcolor = blue]{hyperref}
\usepackage{cite}
\usepackage{upgreek}
\usepackage{multirow}
\usepackage{relsize}

\newcommand \bes {\begin{equation}\begin{split}}
\newcommand \be {\begin{equation}}
\newcommand \ee {\end{equation}}
\newcommand \ees {\end{split}\end{equation}}

\newcommand \dx { \, \mathrm{d}x }
\newcommand \dy { \, \mathrm{d}y }

\newcommand \drv { \, \mathrm{d}\mathbf{r} }

\begin{document}

\title[Explicit temperature coupling in phase-field crystal models]
{Explicit temperature coupling in phase-field crystal models of solidification}

\author{Maik Punke$^1$, Steven M. Wise$^2$, Axel Voigt$^{1,3}$, Marco Salvalaglio$^{1,3}$}
\address{$^1$Institute of Scientific Computing, TU Dresden, 01062 Dresden, Germany.}
\address{$^2$Department of Mathematics, The University of Tennessee, Knoxville, TN 37996, USA.}
\address{$^3$Dresden Center for Computational Materials Science, TU Dresden, 01062 Dresden, Germany.}
\vspace{10pt}

\begin{abstract} 
We present a phase-field crystal (PFC) model for solidification that accounts for thermal transport and a temperature-dependent lattice parameter. Elasticity effects are characterized through the continuous elastic field computed from the microscopic density field. We showcase the model capabilities via selected numerical investigations which focus on the prototypical growth of two-dimensional crystals from the melt, resulting in faceted shapes and dendrites. This work sets the grounds for a comprehensive mesoscale model of solidification including thermal expansion.
\end{abstract}


\section{Introduction}

Solidification of crystalline materials is an ubiquitous phenomenon in nature and technology. It yields shapes and patterns which depend on out-of-equilibrium growth conditions and the interplay of different instabilities \cite{Langer1980}. At the same time, the control of the solidification process is key in several technological applications, from conventional ones \cite{flemings1974solidification,dantzig2016solidification} up to bottom-up approaches exploiting self-assembly \cite{Stangl2004,polshettiwar2009self}. The outcome of crystal growth results from several different contributions: crystal seeds grow after nucleation or on pre-existing crystal phases/seeds, while later stages are affected by capillarity, elasticity, plasticity, as well as various kinetic effects \cite{BOETTINGER200043,hoyt2003atomistic,Jaafar2017,Granasy2019,Alexandrov2021}. When considering solidification, heat transfer in the region surrounding the solid phase also plays an important role \cite{BOETTINGER200043,Jaafar2017,Granasy2019,Alexandrov2021}. From a microscopic point of view, the arrangement of atoms in a periodic lattice typically emerges in anisotropic behaviors such as faceting, as well as in the nucleation and motion of defects, which are tightly related to crystallographic directions. 

The modeling of solidification has been addressed by different approaches. Microscopic approaches are suitable for evaluating lattice-dependent features such as anisotropies and defect structures \cite{hoyt2003atomistic}. However, the growth of crystals involves long time scales, typically not accessible within these methods. Macroscopic approaches, which cope with large systems and long timescales, proved successful in describing the main features of crystal growth via both front tracking and phase-field methods \cite{KOBAYASHI1993410,Karma1998,Zhu2007,Steinbach2009,Pan2010,takaki2014phase,kaiser2020}. However, they usually lack a direct connection to the lattice symmetry and microscopic features in general. Lattice-dependent effects can then be described only partially and are included mainly through parameters and additional functions, e.g. as anisotropic interface energies \cite{SUZUKI2002125,Tor2009}.

The so-called phase-field crystal (PFC) model \cite{Elder2002,Elder2004,Provatas2010,Emmerich2012}
emerged as a prominent approach to describe crystal structures at large (diffusive) timescales through a continuous, periodic order parameter representing the atomic density. It describes solidification and crystal growth, including capillarity, elasticity as well as nucleation and motion of defects. As a result, it reproduces the main phenomenology of crystal growth in two- and three-dimensions \cite{Tegze2011,TANG2011146,Emmerich2012}. Also, it allows for a self-consistent description of anisotropies resulting from the lattice structure \cite{PODMANICZKY2014148,Ofori-Opoku2018}. Therefore, it represents a suitable framework for the development of a comprehensive model of crystal growth. Only recently, however, a temperature field and the related heat flux have been explicitly considered in PFC models \cite{Kocher2019,wang2021thermodynamically}. Similarly, while its first formulation already includes elasticity effects, recent studies shed light on its connections to continuum mechanics \cite{Spatschek2010,HeinonenPRE2014,HeinonenPRL2016,SkaugenPRL2018,SkaugenPRB2018,SalvalaglioNPJ2019,chockalingam2021elastic}, allowing for extended descriptions of the crystal lattice and its deformation.

This paper presents a PFC model of crystal growth accounting for temperature effects through the coupling with thermal transport \cite{wang2021thermodynamically}, and including a temperature-dependent lattice parameter by a suitable modification of the PFC free energy \cite{chockalingam2021elastic}. The resulting model is illustrated in section~\ref{sec:model}. The equations are solved numerically by a simple but efficient approach based on the Fourier pseudo-spectral method as shown in section~\ref{sec:numerics}. Section~\ref{sec:elasticity} reports the description of the continuous elastic field exploited to assess the proposed model. In section~\ref{sec:simulations} the model is then showcased through selected numerical simulations. They reproduce the prototypical growth of two-dimensional crystals from a melt, resulting in faceted shapes and dendrites. The effects of temperature described by the model are illustrated. Finally, we draw conclusions and perspectives in section~\ref{sec:conclusions}.

\section{Model}
\label{sec:model}
The PFC model is based on a Swift–Hohenberg free energy functional \cite{Elder2002,Elder2004,Emmerich2012}, which can be written as
\begin{equation}
F\left[\psi\right] = \int_\Omega \dfrac{\lambda-\kappa(1-q_0^4)}{2}\psi^2 -  \dfrac{\psi^3}{6} + \dfrac{\psi^4}{12}+ \dfrac{\kappa}{2}\left(-2 \nabla (q_0^2\psi) \cdot \nabla \psi + \left(\nabla^2 \psi \right)^2 \right) \drv,   
\label{eq:F1}
\end{equation}
with $\psi\equiv \psi(\mathbf{r},t)$ a scalar order parameter related to the atomic number density, $\Omega \in  \mathbb{R}^n$ its domain of definition ($n=2$ in this work), $q_0$ a length scale enforcing a periodicity of $a_0 \approx 1/q_0$ and a set of material parameters $\lambda,\,\kappa > 0$. Coupled with appropriate boundary- and initial-conditions the time evolution of $\psi$ is described via a conservative
($\mathrm{H}^{-1}$)
gradient flow of $F$, namely
\begin{equation}
\partial _t \psi = \nabla^2 \dfrac{\delta F\left[\psi\right]}{\delta \psi}.
\label{eq:gf1}
\end{equation}
It is worth mentioning that different forms of $F$ are used in literature as, e.g., in the originally proposed formulation \cite{Elder2002} including $q_0$ in the differential operator $\psi(q_0^2+\nabla^2)^2\psi$ or $(q_0^2\psi+\nabla^2\psi)^2$ only, a parameter $\epsilon=\kappa-\lambda$ playing the role of the quenching depth, and eventually additional coefficients for other terms in $\psi$, such as $\tau \psi^3 + \nu \psi^4$ (with $\tau=0$ and $\nu=1$ in \cite{Elder2002}). Such formulations, however, are equivalent and lead to the same expression for $\delta F/\delta \psi$. 
In \cite{wang2021thermodynamically} an extension of this PFC model including heat-transfer through a temperature field was proposed. Therein, the (dimensionless) Helmholtz free energy functional with $q_0=1$, here normalized with respect to the temperature, reads 
\begin{equation}
    \begin{split}
\mathcal{F}\left[\psi, T\right] & := 
\frac{F\left[\psi, T\right]}{T}
    \\
& = \int_\Omega \dfrac{\lambda }{2}\psi^2 -  \dfrac{ \psi^3}{6} + \dfrac{ \psi^4}{12}  - \vartheta \mathrm{ln}(T) - \frac{1}{T}\gamma (\psi +1)+ \dfrac{\kappa  }{2}\left(-2\left| \nabla\psi \right|^2 + \left(\nabla^2 \psi \right)^2 \right) \drv,
\label{eq:F2}
    \end{split}
\end{equation}
with $T\equiv T(\mathbf{r},t)$ the dimensionless temperature and $T=1$ at the melting point. $\vartheta, \gamma >0$ are additional parameters.  The dynamics is then given by the coupled evolution of $T$ and $\psi$, reading 
\begin{equation}
\begin{split}
   \gamma \partial_t T- \vartheta \partial _t \psi    &= M \nabla^2 T, \\
    \partial_t \psi &= \nabla^2  \dfrac{\delta \mathcal{F}\left[\psi,T\right]}{\delta \psi},
    \end{split}
    \label{eq:psiEvol}
\end{equation}
with $M>0$ a parameter corresponding to the thermal diffusivity, assumed to be constant. 

This model is here extended by including a variable lattice parameter and connecting it to the temperature field, aiming at describing thermal expansion in the crystal phase. Let us consider a crystalline solid with lattice parameter $a_0$. Expansion due to temperature can be described through the thermal expansion coefficient $\alpha^{\rm th}=\bar{L}^{-1}({\rm d} L/{\rm d} T)$ with $\bar{L}$ a characteristic (reference) length of measurement and $L$ a linear dimension in the solid. Assuming weak dependence of $\alpha^{\rm th}$ on $T$, $a_{T}=a_0[1+\alpha^{\rm th}(T-T_0)]$ with $a_T$ and $a_0$ the lattice spacing at $T$ and $T_0$, respectively. This equation defines a relative deformation, i.e. a thermal strain $\varepsilon^{\rm th}(T,T_0)=(a_T-a_0)/a_0=\alpha^{\rm th}(T-T_0)$.  Following the arguments in \cite{chockalingam2021elastic}, this information may be encoded in the PFC free energy by noticing that $(a_T-a_0)/a_0=(q_0/q_T)-1$ and thus $q_T=\beta(T,T_0) q_0$ with $\beta \equiv \beta(T,T_0)=1/(1+\varepsilon^{\rm th}(T,T_0))$. Therefore, by setting $q_0=1$ we may derive from equations~\eqref{eq:F1} and \eqref{eq:F2} the following free energy functional:
\begin{equation}
\label{eq:freeE}
\begin{split}
\mathcal{F}_\beta\left[\psi, T\right] := 
\frac{F_\beta\left[\psi, T\right]}{T}  =  \int_\Omega &  \dfrac{\lambda - \kappa (1-\beta ^4)}{2} \psi^2 -  \dfrac{ \psi^3}{6} + \dfrac{ \psi^4}{12}  - \vartheta  \mathrm{ln}(T) - \frac{1}{T}\gamma (\psi +1) 
    \\
 &+ \dfrac{\kappa }{2}\left(-2 \nabla\psi \cdot \nabla (\beta^2 \psi) + \left(\nabla^2 \psi \right)^2 \right) \drv.
\end{split}
\end{equation}
This leads to the following evolution equations defining our model:
\begin{equation}
\label{eq:evol}
\begin{split}
& \gamma \partial_t T- 
\vartheta \partial _t \psi  = M \nabla^2 T,\\
& \partial_t \psi =  \nabla^2 \left[(\lambda - \kappa (1-\beta ^4)) \psi -  \dfrac{ \psi^2}{2} + \dfrac{ \psi^3}{3}  + \kappa \beta^2 \nabla^2\psi + \kappa \nabla^2(\beta^2 \psi) + \kappa \nabla ^4 \psi - \dfrac{\gamma}{T} \right].
\end{split}
\end{equation}
Note that the PFC heat-flux model, equations~\eqref{eq:F2}-\eqref{eq:psiEvol}, without temperature-dependent lattice parameter \cite{wang2021thermodynamically} is recovered for $\beta = 1$. On the other hand, by neglecting the heat flux with a constant and unitary temperature field, and setting a constant $\varepsilon^{\rm th}$, the evolution equation for the density $\psi$ in the presence of an eigenstrain $\varepsilon^*=\varepsilon^{\rm th}$ as introduced in reference~\cite{chockalingam2021elastic} is obtained. Therefore, our resulting model, equation~\eqref{eq:evol}, realizes a general combination of these two models.

In this work we consider honeycomb and triangular two-dimensional crystals. In a one-mode approximation, they are well described by
\begin{equation}
    \psi(\mathbf{r},t) \approx \overline{\psi}(\mathbf{r},t) + \sum_{n=1}^N \eta_n (\mathbf{r},t) {\rm e}^{\mathbb{i} \mathbf{q}^n \cdot \mathbf{r}},
    \label{eq:ansatz}
\end{equation} 
with $\mathbb{i}$ the imaginary unit, $\overline{\psi}\equiv \overline{\psi}(\mathbf{r},t)$ the local average density, $\eta_n\equiv \eta_n(\mathbf{r},t)$ amplitudes function which are real for relaxed crystals, $\eta_n= \phi$, and reciprocal lattice vectors ($N=6$), $\lbrace \mathbf{q}^n\rbrace_{n=1}^6= q_0 \lbrace (0,\,\pm 1),\,(\pm \sqrt{3}/2, \pm1/2)\rbrace$. According to energy minimization, the real amplitude in the relaxed solid phase results
\begin{equation}
\label{eq:crystalAnsatz}
\phi \equiv \phi_{\pm} =\dfrac{2-4A \pm 2\sqrt{1+16A+20(\kappa-\lambda-16A^2)}}{24},
\end{equation}
with $A$ the global average density. For $A>0.5$ $(A<0.5)$ the amplitudes can be identified with $\phi_{-}$ $(\phi_{+})$ resulting in honeycomb and triangular structure, respectively (with similar energetics). 
Equation \eqref{eq:crystalAnsatz} is exploited to initialize $\psi$ in the solid phase, whereas the density of the pure liquid phase is initialized as $\overline{\psi} = A$ and $\phi=0$. Also the temperature field is initialized as spatially homogeneous $T=T_0$. Parameters used in the simulations are reported in table \ref{tab:numerics}.

\section{Numerical methods}
\label{sec:numerics}

\begin{figure}[t]
	\includegraphics{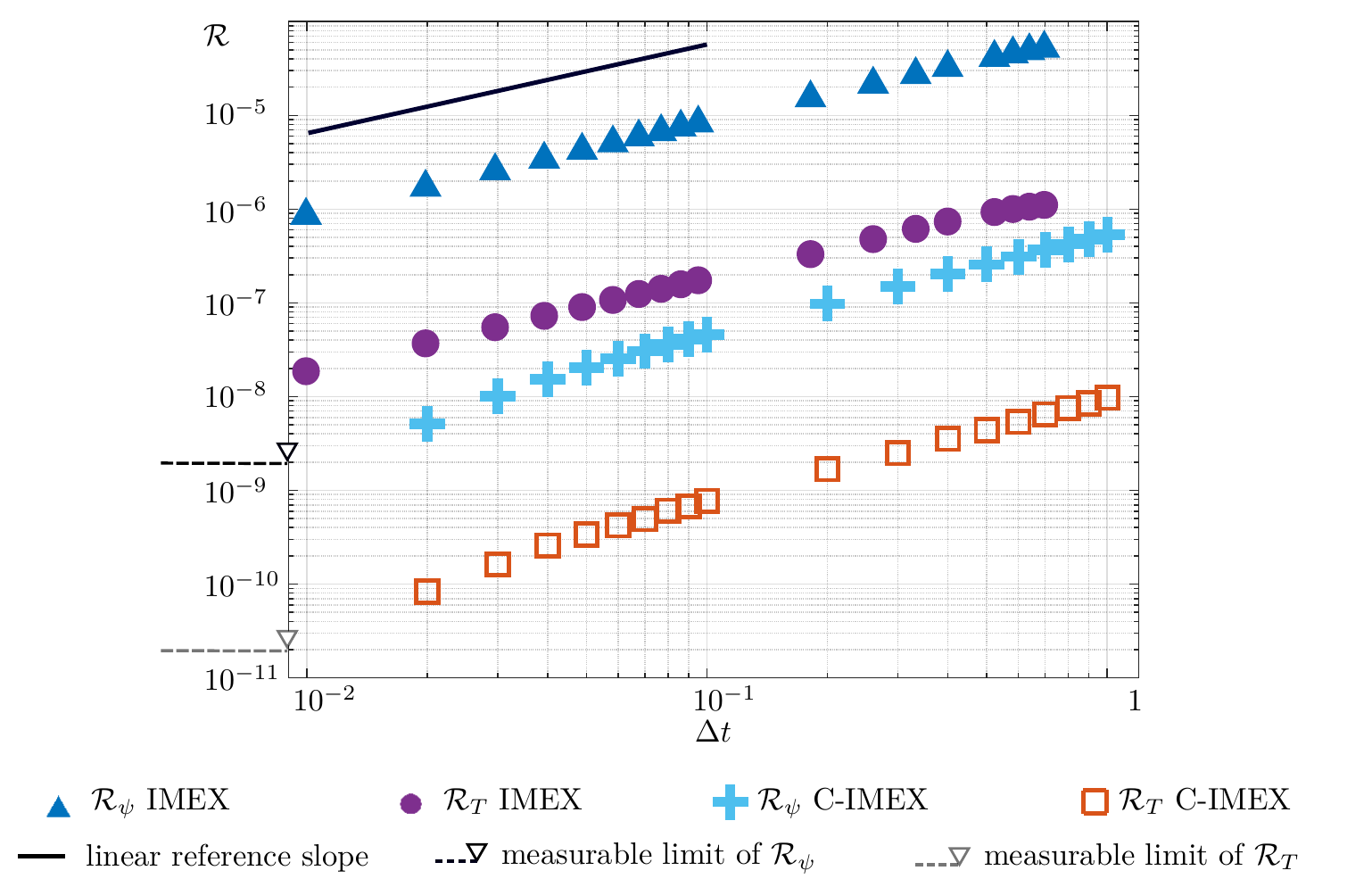}
	\caption{Convergence study of a C-IMEX scheme (with $C=-0.149$) compared to a standard IMEX scheme ($C=0$). The measurable limit of the accuracy $\mathcal{R}$ is obtained by a linear extrapolation. 
	}
	\label{fig:convStudy}
\end{figure}

\begin{table}[t]
    \centering
    \resizebox{\textwidth}{!}{
    \begin{tabular}{|c||ccc|ccc||ccc|}
    \hline
     Figure& \multicolumn{6}{c||}{Model Parameters}&\multicolumn{3}{c|}{Numerical Parameters} \\
         &$\lambda$ &$\kappa$  &A& $\gamma$ &$T_0$ & $\alpha^\mathrm{th}$ &$C$&$K$&$\Delta t$ \\
         \hline \hline 
         \ref{fig:eshelby}&$1.04$ &$1$  & $0$ &$0$&-&0&-0.149&0.990&5\\
         \hline
          \multirow{2}{*}{\ref{fig:longTimeSim}}&$0.6$ &$0.45981$  & $0.849$ &$ 0.06$&0.6&&&&\\
          & &  &  &&&1.0&-1.060&1.000&1\\
          \hline
          \multirow{6}{*}{\ref{fig:parameterStudy}(a)}&&&&&0.1&0&0.100&&0.5\\
            & & & &&$0.2 \dots 0.3$&&-0.140&0.998&3\\
                         & & & &&0.4&&-0.147&1.000&5\\
              & & & &&0.5&&&0.990&\\
               & & & &&0.6&&-0.149&0.990&\\
              &&&&&$0.7\dots 0.97$&&&0.985&\\
               \hline
    \multirow{11}{*}{\ref{fig:parameterStudy}(b)}
      &&&&$0$&-&&&0.990&\\
    &&&&$0.01\dots 0.1$&0.6&&&&\\
          & & &&$0.2$&&&-0.143&1.020&\\
          & & & &$0.3\dots 0.6$&&&-0.130&1.022&3\\
                & & & &$0.7\dots 0.9$&&&0.050&1.000&1\\
           & &  &&$1$&&&$0.200$&&0.5\\
           & &  &&$2$&&&$1.000$&$1.030$&0.4\\
            & &  &&$3$&&&$2.100$&$1.050$&\\
             & &  &&$4$&&&&&0.3\\
              & &  &&$5$&&&&&0.2\\
               & &  &&$6$&&&&&0.1\\
               \hline
        \multirow{2}{*}{\ref{fig:parameterStudy}(c)}     & & &&0.06&&$5\cdot 10^{-4}\dots 0.5$&-1.060&0.995&1\\     & & &&&&$0.6\dots 1.3$&&1.000&\\ 
                    \hline
        \multirow{3}{*}{\ref{fig:wulff}(a)} 
        & &  &$0.844,\,0.846$&&&0&-0.149&0.974&5\\
             &  & &0.849&&&&&0.990&\\
             & 0.2&$1/3$  &1.120&&&&0.130&1.000&2\\
             \hline
        \ref{fig:wulff}(b),\,(c) 
        & 0.6&0.45981  &$0.849$&&&&-0.149&0.990&5\\
        \hline
         \multirow{2}{*}{\ref{fig:merging},\,\ref{fig:stress4},\,\ref{fig:stress3}}
         & &  & $0.849,\, 0.151$ &&&&&&\\
          & &  &  &&&1.0&-1.060&1.000&1\\
                                 \hline   \end{tabular}}
    \caption{Numerical parameters $(C,K,\Delta t)$ for all the simulations reported in this paper. We set $M/\gamma = \vartheta/\gamma = 1$ except for $\gamma = 0$ (decoupled equations in \eqref{eq:evol}), for which we set $M=\vartheta = 0$. Empty table entries read as the row above.}
    \label{tab:numerics}
\end{table}

Numerical solutions of the partial differential equations \eqref{eq:evol} are computed using a first-order linear IMEX scheme for the time discretization in combination with a Fourier pseudo-spectral method for the spatial discretization, enforcing periodic boundary conditions. The numerical simulations are performed on a uniform grid with an element size of $\dx = \dy = 0.78125$, resulting in a resolution of $\approx 75$ grid points per unit cell. The Fourier transforms are performed with an algorithm based on the FFTW2 library. 
A constant time step size $\Delta t$ for the IMEX scheme is considered, and chosen to ensure numerical convergence. Additionally, we exploit the numerical time-stabilization routine presented in \cite{elsey2013simple}. This approach features a convex-concave splitting of the free energy controlled by a parameter $C$, combined with an IMEX scheme (C-IMEX). For every set of model parameters, we determine optimal $C$ by minimizing the difference of the error (least square of the difference) in the free energy decay obtained with the C-IMEX scheme with respect to a numerical reference solution (IMEX scheme with small timestep). Optimal errors are found by considering a minor rescaling of the time scale by a factor $K\approx 1$, which is found to compensate for quantitative effects on the time scale observed for the C-IMEX scheme \cite{elsey2013simple}. 

A convergence study of the considered numerical methods is reported in figure~\ref{fig:convStudy}, for a specific set of model parameters leading to the growth of a crystal in domain $\Omega= [-28p, 28p]^2$ and for time steps $\Delta t \in [10^{-2}, 1]$. This figure shows the residua of $\psi$ and $T$, $\mathcal{R}_\psi$ and $\mathcal{R}_T$, respectively, evaluated as the integral over the domain of the squared difference from a numerical reference solution for different $\Delta t$. The IMEX ($C=0$) and C-IMEX ($C=-0.149$) schemes are compared, with a reference solution corresponding to the C-IMEX scheme with $\Delta t =0.01$. As expected, the schemes converge linearly for a decreasing time step size $\Delta t$. However, for a fixed $\Delta t$, the C-IMEX approach gives two orders of magnitude smaller error than the IMEX approach for $\psi$ and a two and a half orders of magnitude smaller error for $T$. By fixing an accuracy instead, the C-IMEX approach allows for two-order of magnitude larger timesteps compared to the IMEX approach. Convergence studies similar to figure~\ref{fig:convStudy} have been performed for all the simulations reported in this work. Table \ref{tab:numerics} summarizes the numerical parameters $(C,K,\Delta t)$, together with other model parameters for all simulations in this paper.

\section{Evaluation of elastic fields}
\label{sec:elasticity}

In this section we address the evaluation of the elastic field to assess the PFC formulation including a variable lattice spacing as introduced in section~\ref{sec:model}. Elasticity in PFC models can be characterized by small perturbations of the equilibrium density $\psi$. Therefore, it features variations of $\psi$ over a length scale significantly larger than the lattice spacing, and a coarse-graining of $\psi$ is usually considered in this context \cite{SalvalaglioNPJ2019,skogvoll2021stress}.  
In particular, one may consider an expansion of the periodic density as in equation~\eqref{eq:ansatz} with complex amplitudes encoding lattice distortion over a displacement field $\mathbf{u}$. Indeed, setting $\psi(\mathbf{r},t)=\psi^{\rm ref}(\mathbf{r}-\mathbf{u},t)$ with reference density field $\psi^{\rm ref}(\mathbf{r},t)$ leads to equation~$\eqref{eq:ansatz}$ with $\eta_n (\mathbf{r},t)=\phi(\mathbf{r},t)\mathrm{e}^{-\mathbb{i} \mathbf{q}^n \cdot \mathbf{u}(\mathbf{r})}$. 
This ansatz can be also exploited to derive a coarse-grained PFC model, namely the amplitude expansion of the PFC (APFC), where $\eta_n$ are the variables to solve for \cite{Goldenfeld2005,Athreya2006,salvalaglio2022coarse}. Within the PFC model, $\eta_n$ as well as $\overline{\psi}$ can be computed from a demodulation of $\psi$ \cite{skogvoll2021stress,SkogvollJMPS2022}:
\begin{equation}
\label{eq:amplitudeNew}
\begin{split}
    \eta_n &= \mathrm{e}^{-\mathbb{i}  q_j^n r_j}\mathscr{F}^{-1}\left[\mathrm{e}^{-2\pi a_j^2 (k_j-q^n_j)^2 }\mathscr{F}\left[ \psi \right] \right],\\
\overline{\psi} &= \mathscr{F}^{-1}\left[\mathrm{e}^{-2\pi a_j^2 k_j^2 }\mathscr{F}\left[ \psi \right] \right].
    \end{split}
    \end{equation}
with $\mathscr{F}$ the Fourier transform, $\mathscr{F}^{(-1)}$ the inverse Fourier transform, and implying Einstein summation convention. The demodulation \eqref{eq:amplitudeNew} includes a coarse-graining procedure over one unit cell $\mathrm{UC}=[0, a_1]\times [0, a_2]$ with $a_1 = p$, $a_2 = \sqrt{3}p/2$ and  $p=4\pi/\sqrt{3}$.
Following recent works \cite{SkaugenPRL2018,skogvoll2021stress,SalvalaglioJMPS2020,chockalingam2021elastic}, one may obtain the mechanical stress tensor $\sigma_{ij}\equiv \sigma_{ij}(\mathbf{r},t)$ without isotropic pressure terms from the density field:
\be
\label{eq:stress1}
\begin{split}
\sigma_{ij} &= \Big\langle\dfrac{\delta (\mathcal{F}_\beta [\psi(\mathbf{r}+\mathbf{u},t), T(\mathbf{r},t)]-\mathcal{F}_\beta [\psi(\mathbf{r},t), T(\mathbf{r},t)])}{\delta \partial_{i}u_j}\Big \rangle\\
&= \kappa \Big\langle \left(2 \beta^2 \psi_{,i}+\psi_{,kki}+ \psi (\beta^2)_{,i}      \right)\psi_{,j} - \psi_{,kk} \psi_{,ij}  \Big\rangle.
\end{split}
\ee
with $\langle \cdot \rangle$ a spatial average over the unit cell, which may be explicitly performed in reciprocal space through the smoothing kernel similarly to equation~\eqref{eq:amplitudeNew} for $\overline{\psi}$. Equation~\eqref{eq:stress1} generally defines $\sigma_{ij}$ up to a phase-dependent divergence-free contribution, which can be removed by considering $\psi$ as expressed through an one-mode approximation as in equation~\eqref{eq:ansatz} \cite{skogvoll2021stress}. A coarse-grained description of the crystal is also achieved through amplitudes from equation~\eqref{eq:amplitudeNew}. 

\begin{figure}[t]
	\centering
	\includegraphics[width=0.85\textwidth]{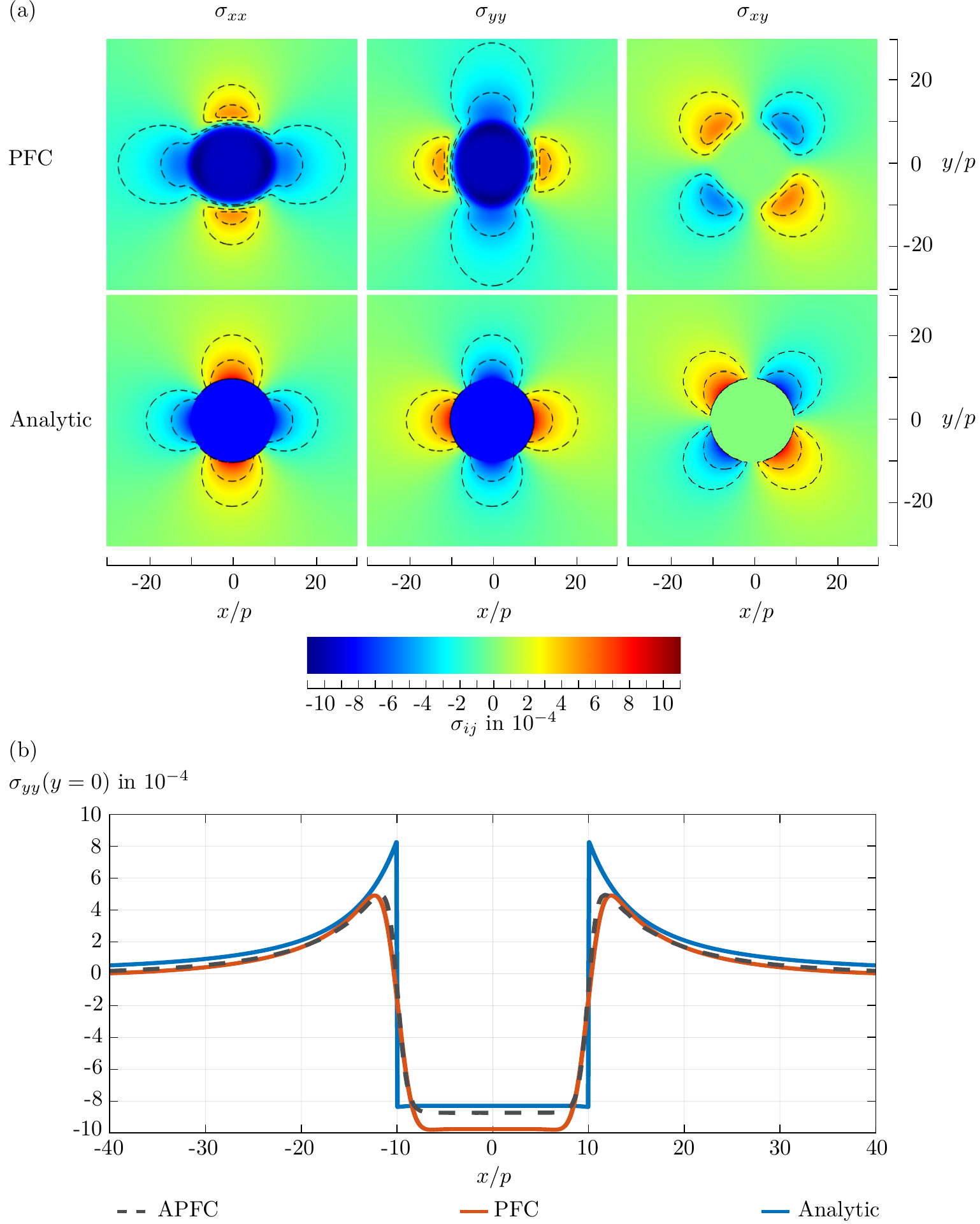}
	\caption{Benchmark of the stress definition, equation~\eqref{eq:stress1}, through the Eshelby inclusion problem. (a) Stress field components extracted from the result of a PFC simulation (first row) and the analytic solution for an inclusion in an infinite medium (second row). (b) Comparison of the $\sigma_{yy}(x)$ for these solutions and the result of corresponding APFC calculations from reference~\cite{chockalingam2021elastic} at $y$ corresponding to the center of the inclusion. Lengths are scaled with the atomic spacing along the $x$ direction, $p$.}
	\label{fig:eshelby}
\end{figure}
Assuming $\eta_n$ to be constant over the length of UC (similarly to the basic assumption of the APFC model \cite{salvalaglio2022coarse}) and denoting the complex conjugate of $ \eta_{n}$ as $\eta^*_{n}$, we obtain
\begin{equation}
\label{eq:stress}
\begin{split}
\sigma_{ij} =\kappa \sum_{n=1}^N \bigg[ & \Big(2\langle\beta^2 \rangle(\partial_i + \mathbb{i}q^n_{i}) \eta_n +  (\partial_i +\mathbb{i}q^n_{i})(\partial_{kk} + 2\mathbb{i} q^n_k \partial_k -q^n_k q^n_k)\eta_n + \eta_n \langle\beta^2\rangle_{,i}\Big)   \\
&(\partial_j -\mathbb{i}q^n_{j}) \eta^*_{n} -\left(\partial_{kk} + 2\mathbb{i} q^n_k \partial_k- q^n_k q^n_k\right) \eta_n (\partial_i - \mathbb{i}q^n_{i}) (\partial_{j} -\mathbb{i}q^n_{j})  \eta^*_{n}\bigg].
\end{split}
\end{equation}
Equation~\eqref{eq:stress1} with an one mode approximation for $\psi$ and equation~\eqref{eq:stress} are found to deliver almost identical elastic fields. As will be shown below, this approach allows to characterize elastic properties also in systems featuring solid-liquid coexistence.

In reference \cite{chockalingam2021elastic} a benchmark problem has been reported for the APFC model. It consists of the simulation of an elastic inclusion through an eigenstrain formulation that enforces a spatially-dependent lattice parameter through a quantity similar to $\beta$ in equation~\eqref{eq:freeE}. It corresponds to the so-called Eshelby problem  \cite{eshelby1957determination,eshelby1959elastic,mura2013micromechanics}. Here we consider the same setting to assess this elastic inclusion problem within the microscopic PFC model (rather than the coarse-grained APFC model). We set a spatially dependent, but constant in time, lattice parameter
\begin{equation}
\label{eq:beta}
\begin{split}
    \beta = 1-\left(1-\dfrac{1}{1+\varepsilon^*}\right)\xi, \qquad
    \xi=\dfrac{1}{2}\left(1-\mathrm{tanh}\left(\dfrac{|\mathbf{r}|-R}{\varepsilon} \right)\right),
\end{split}
    \end{equation}
with $\varepsilon^* =0.01,\, R = 10p,\,\varepsilon = p$. The domain $\Omega$ is set to $ [-50p,50p]\times[-25\sqrt{3}p,25 \sqrt{3}p]$ for a triangular lattice and we let the system relax until a steady state is reached.
Figure~\ref{fig:eshelby} provides a comparison of the mechanical stress components computed from the PFC simulation through equation~\eqref{eq:stress} and the analytic solution for an inclusion in an infinite medium \cite{mura2013micromechanics,li2005circular}. For comparison, the result obtained by an APFC simulation \cite{chockalingam2021elastic} is also reported. A good agreement between the analytic solution and the simulations is obtained. A few differences may be recognized and ascribed to different effects similarly to reference~\cite{chockalingam2021elastic}. In brief, the considered setting models the inclusion problem inside a domain of finite size and with periodic boundary conditions, while the Eshelby solution is derived for infinite systems. Also, elastic properties of PFC models account for non-linear effects, which are absent in the analytic solution. Furthermore, the approximation of the inclusion's boundary as a diffuse interface achieved through equation~\eqref{eq:beta} inherently produces a smooth field that differs from the sharp transition implied in the analytic solution. This smoothing and the decay of the elastic field away from the inclusion match the results obtained with the APFC model with the same parameters. A slightly different stress field in the inclusion is obtained, that can be ascribed to a fixed average density assumption in the considered APFC simulation.

\begin{figure}[t]
	\includegraphics{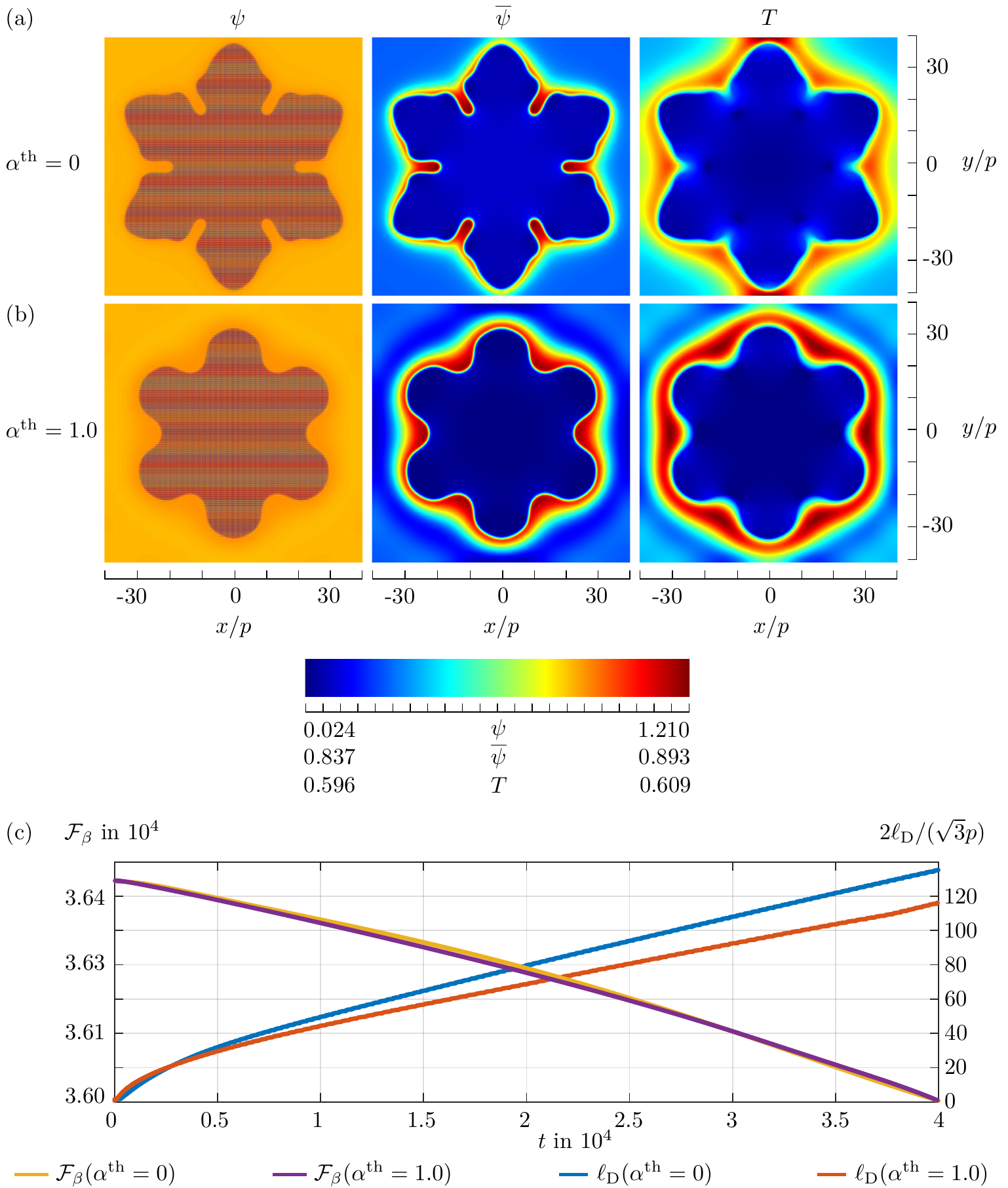}
	\caption{Numerical solution of equation~\eqref{eq:evol} for a given set of parameters, with and without including a temperature-dependent lattice spacing. The density field $\psi$, local average density $\overline{\psi}$ and temperature field $T$ during growth of the dendrite at $t=3.5\cdot 10^4$ are shown for (a) $\alpha^\mathrm{th}=0$ and (b) $\alpha^\mathrm{th}=1.0$. (c) time evolution of the free energy $\mathcal{F}_\beta$ and length of dendritic arms $\ell_{\mathrm{D}}$ (scaled by the length of the unit cell along the y-direction) for $\alpha^\mathrm{th}=0$ and $\alpha^\mathrm{th}=1.0$. $\Omega=[-110p,110p]^2$.
	}
	\label{fig:longTimeSim}
\end{figure}

\section{Numerical Simulations}
\label{sec:simulations}
\subsection{Model features and parameter study}
\label{subsec:modelProperties}
In this section, we illustrate the capabilities of the proposed model through numerical experiments exploiting the method described in section~\ref{sec:numerics}. In particular, we provide an overview of its features focusing on the dendritic solidification regime. 
Figure~\ref{fig:longTimeSim}(a)--(b) illustrates the numerical solution of equation~\eqref{eq:evol} at a representative stage ($t=3.5\cdot 10^4$) for $A=0.849$ ($A>0.5$, honeycomb), with and without including a temperature-dependent lattice spacing (namely with $\alpha^\mathrm{th}=0$ and $\alpha^\mathrm{th}=1.0$).  Focusing first on the morphologies of the crystalline domain, we note that the initial crystal seed grows in both cases and develops an anisotropic shape. It corresponds to a six-fold dendrite which reflects the underlying triangular symmetry of the crystal lattice. This evidence extends the preliminary numerical results reported in reference~\cite{wang2021thermodynamically}, showing that the model including heat flux allows for dendritic growth with an extended set of parameters. Moreover, differences are observed among the simulations with $\alpha^\mathrm{th}=0$ and $\alpha^\mathrm{th}=1.0$. The latter case leads to growing arms of the dendrite, which are more rounded than the former. As shown in figure~\ref{fig:longTimeSim}(c) reporting a comparison of the length of dendritic arms over time $\ell_{\mathrm{D}}\equiv \ell_{\mathrm{D}}(t)$ in both cases, the velocity of the tip is also smaller for $\alpha^\mathrm{th}=1.0$ (namely $\ell_{\mathrm{D}}/t\approx 1.85\cdot 10^{-2}$ and $\ell_{\mathrm{D}}/t\approx 1.50\cdot 10^{-2}$ for $\alpha^\mathrm{th}=0$ and $\alpha^\mathrm{th}=1.0$, respectively). However, in both cases, after a first initial transient, a constant tip velocity is observed, consistently with a classical result for dendritic growth \cite{Alexandrov2021}.  Figure~\ref{fig:longTimeSim}(c) also includes the time evolution of $\mathcal{F}_\beta$, showing that the energy decreases for both the considered settings.

For the chosen parameters, the growing crystal has a local average density lower than the liquid phase and, as dictated by the equation for $\partial_t T$, a lower temperature with respect to $T_0$ (see figure~\ref{fig:longTimeSim}(a)--(b)). This behavior is observed, e.g., for materials such as hexagonal ice which is well represented by a honeycomb structure, but also diamond-cubic silicon, rhombohedral bismuth, and orthorhombic gallium \cite{OHRING1995189}.
At the liquid-solid interface, the temperature increases above $T_0$, wheres away from the growing seed where $\psi=A$, we have $T\approx T_0$. These variations can be interpreted in terms of a Gibbs-Thompson effect \cite{alexandrov2020gibbs,tiller1991science}, commonly observed at phase boundaries, combined with the conservation laws. A region with a slight material depletion surrounding the $T>T_0$ region is observed for $\alpha^{\rm}=1.0$, which can be ascribed to a  material transfer towards the interface triggered by the reduced lattice spacing in the solid phase. In turn, larger temperature gradients are present, which enforce the more isotropic shapes. The more typical behavior of larger (local) average density and $T>T_0$ in the crystal phase can be straightforwardly obtained by setting a global average density $A\mapsto 1-A$ (i.e., for the corresponding triangular phase, as can be readily shown from the equations defining the model). These cases lead to identical energetics and, in turn, morphologies and fields for the $\alpha^\mathrm{th}=0$ case. However, for $\alpha^\mathrm{th}=1.0$, either a lattice expansion or compression is enforced for settings with symmetric global average densities with respect to $A=0.5$. As a result, differences are observed: for $A<0.5$ no slight material depletion is observed from the liquid phase with instead a larger region with an increased average density, temperature gradients closely resembling the $\alpha^\mathrm{th}=0$ case and, in turn, similar morphologies (not shown), in agreement with the arguments reported above.
This symmetry breaking will be exploited in the following section to characterize the thermal stress.
\begin{figure}[t]
	\includegraphics{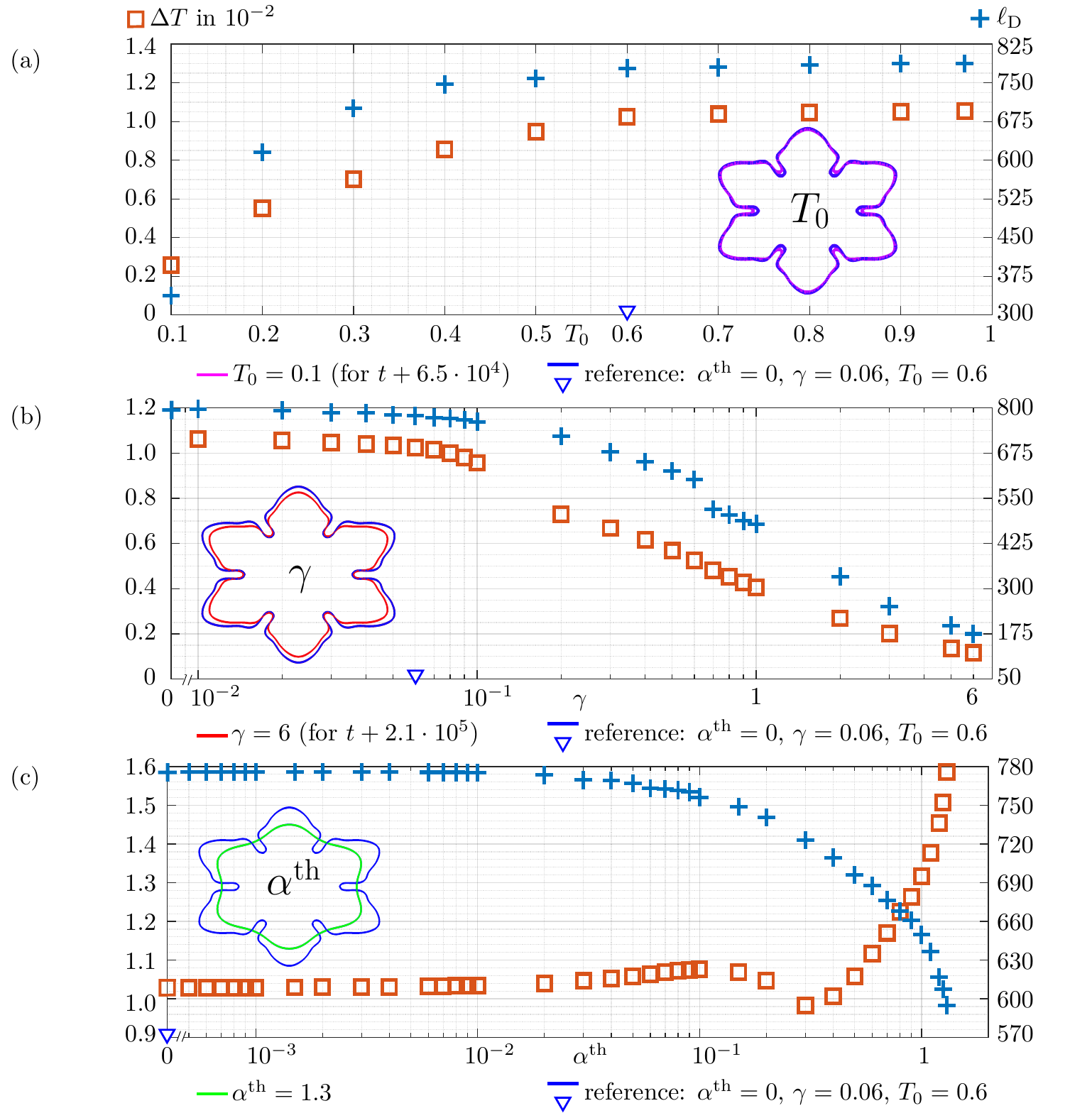}
	\caption{Parameter study of equation~\eqref{eq:evol} with parameter ranges as in \eqref{eq:parrange}. The effects of (a) $T_0$, (b) $\gamma$, and (c) $\alpha^\mathrm{th}$ are illustrated. Insets show representative comparisons of morphologies with a reference (corresponding to the empty triangle in every panel).
	}
	\label{fig:parameterStudy}
\end{figure}

\begin{figure}[t]
	\includegraphics{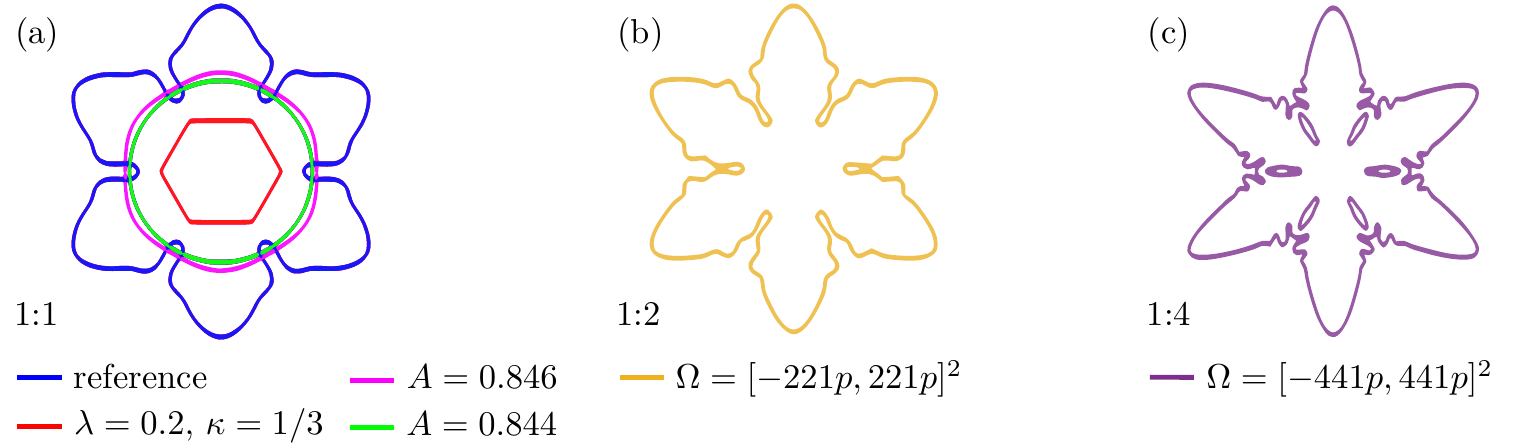}
	\caption{Morphologies obtained for different parameters entering the classical PFC model and the domain size. (a) Effect of different average densities and material parameters (reference shapes as in figure~\ref{fig:longTimeSim}). (b) Shape of a dendrite in a domain $\Omega = [-221p,221p]^2$ scaled by factor 1/2. (c) $\Omega=[-441p,441p]^2$ scaled by factor 1/4.}
	\label{fig:wulff}
\end{figure}

More insights into the morphological evolution are obtained through a parameter study for the same simulation setup considered above. In particular, to shed light on temperature-dependent effects, we vary $ T_0$, $\gamma$, $\alpha^\mathrm{th}$ separately in the following range:
\begin{equation}
        T_0  \in [ 0.1,0.97], \quad       \gamma \in [0, 6] ,\quad \alpha^\mathrm{th}\in [0, 1.3].
        \label{eq:parrange}
\end{equation}
We analyze $\ell_{\mathrm{D}}$ as well as the temperature range, $\Delta T(t)\equiv\max_{\mathbf{x}\in \Omega}T(\mathbf{x},t)-\min_{\mathbf{x}\in \Omega}T(\mathbf{x},t)$, and the morphologies at a given time ($t=3.5\cdot 10^4$), as illustrated in figure~\ref{fig:parameterStudy}. 
Deacreasing $T_0$ or increasing $\gamma$ leads to a decrease in $\Delta T$ and $\ell_{\mathrm{D}}$ at a given time, and vice versa. Also, small changes in the shapes are obtained as illustrated in insets of
figure~\ref{fig:parameterStudy}(a)--(b) comparing representative morphologies. We conclude that the initial temperature, $T_0$, and the coupling strength between equations~\eqref{eq:evol}, $\gamma$, mainly affect the growth rate. In contrast, a large thermal expansion coefficient, $\alpha^\mathrm{th}$, leads to more significant differences in terms of morphologies consistently with the result in figure~\ref{fig:longTimeSim}, as well as a decrease of $\ell_{\mathrm{D}}$ and increase of $\Delta T$. Still, the growth dynamic results mostly unaffected for $\alpha^\mathrm{th}<0.1$.
While kept fixed in the simulations discussed above, parameters entering the original PFC model, like $A$, $\lambda$, $\kappa$, as well as the size of the simulation domain, still determine the features of the solidification process and if the growth leads to a dendritic shape. For the sake of completeness, we illustrate in figure~\ref{fig:wulff} the main expected changes for the solidification process. Varying the average density affects the anisotropy of the growing crystal/dendrite. Also, effectively changing the quenching depth through $\lambda$ and $\kappa$ may lead to a different growth regime, up to reaching equilibrium with a morphology corresponding to the anisotropic equilibrium crystal shape, see figure~\ref{fig:wulff}(a). It is worth mentioning that the growth of dendrites discussed so far may be considered in a larger domain allowing for the development of long arms and the onset of the growth of secondary arms. This is illustrated in figure~\ref{fig:wulff}(a)--(c).

\subsection{Crystal growth and lattice deformation}
\label{subsec:elasticProp}

\begin{figure}[h!]
	\centering
	\includegraphics{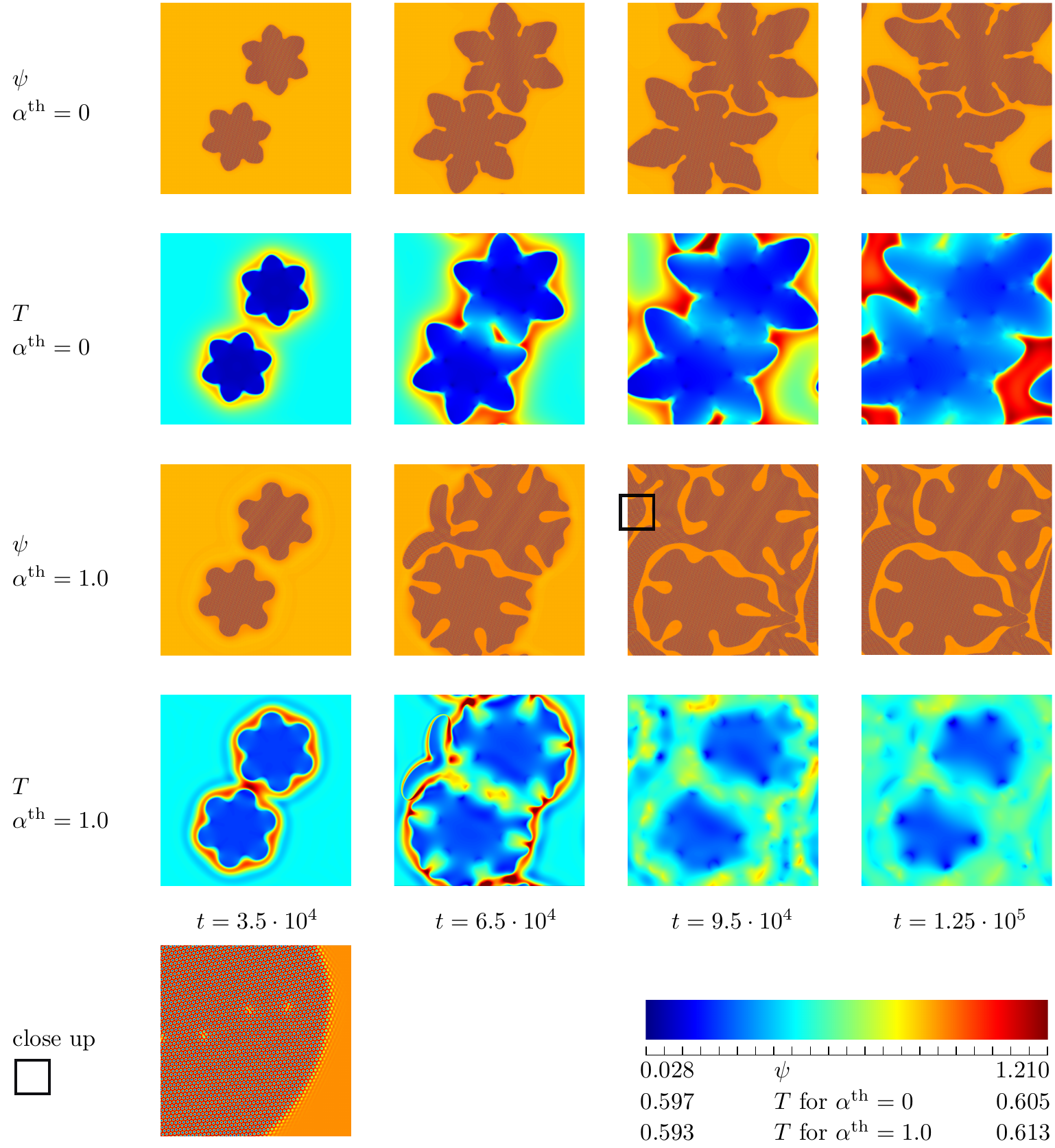}
	\caption{Growth and interaction of two dendrites, with ($\alpha^\mathrm{th}=1.0$) and without ($\alpha^\mathrm{th}=0$) a temperature dependent lattice spacing. The density- and temperature fields at representative stages are reported. Last panel on the bottom reports a magnification of the density field for the $\alpha^\mathrm{th}=1.0$ simulation, explicitly showing defected merging region. Parameters are set as in figure~\ref{fig:longTimeSim}, $\Omega =[-221p,221p]^2$.}
	\label{fig:merging}
\end{figure}

After characterizing the model, we study the growth in a prototypical case representative of more general settings. 
We consider the solidification occurring in a system with two crystal seeds having different (random) orientations. We address both the cases with ($\alpha^\mathrm{th}=1.0$) and without ($\alpha^\mathrm{th}=0$) a temperature-dependent lattice spacing. Figure~\ref{fig:merging} shows the resulting density- and temperature fields at four representative stages. The dendrites exhibit shapes analogous to figure~\ref{fig:longTimeSim} owing to the same parameters. Moreover, when the length of the arms approaches half the distance among the seeds, the temperature fronts where $T>T_0$ meet, and the two growing crystals begin to interact. At this stage, similarly to results reported in the literature \cite{kaiser2020,chen2021phase}, the growth rate drops. Interestingly, while for longer times, the dendritic arms are still separated in case $\alpha^\mathrm{th}=0$, merging is observed for the $\alpha^\mathrm{th}=1.0$ case, with the formation of grain boundaries and dislocations. In general, the merging of growing crystals depends on the phase of the periodic lattice and its deformation at the corresponding growth fronts. If they would form a relaxed crystal, i.e. if the crystal structures meeting at the growth fronts are commensurate, the process is expected to occur. Otherwise, additional energy barriers exist, which may lead to the hindering of the merging process. The evidence from figure~\ref{fig:merging} suggests that the effects induced by the temperature dependence of the lattice spacing in the specific setup considered for this simulation favor the commensurability of the crystals at the growth fronts. 

\begin{figure}[t]
	\includegraphics{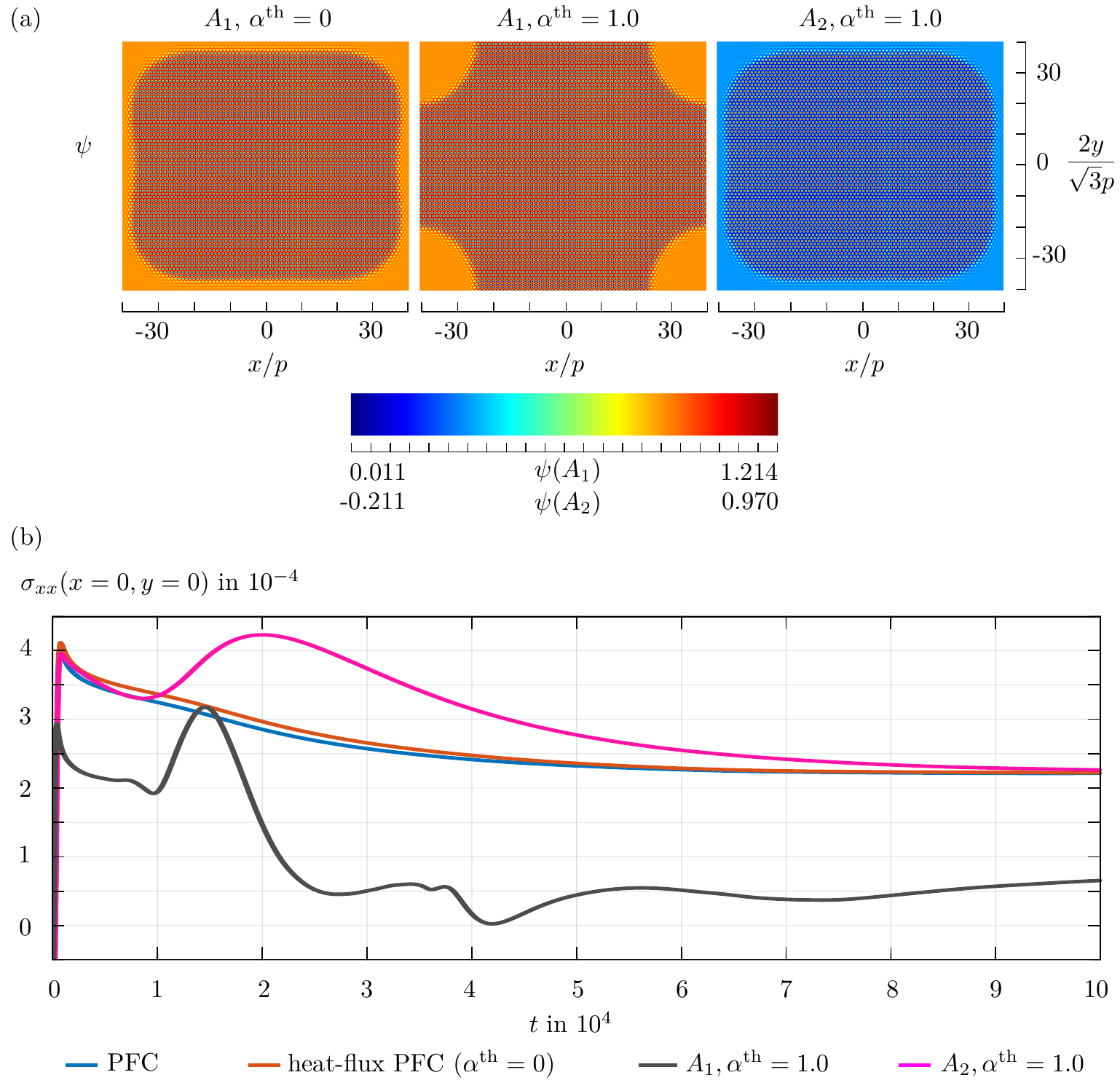}
	\caption{Merging of a small crystal with its periodic images. (a) density $\psi$ for the heat-flux PFC model without  ($\alpha^\mathrm{th}=0$) and with ($\alpha^\mathrm{th}=1.0$) a temperature dependent lattice in cases $A_1=0.849$ and $A_2=0.151$ after $t=10^5$ are shown. (b) time evolution of $\sigma_{xx}$ at the center of the crystal, also reported for the classical PFC model ($T_0=1.0$,$\gamma=0$). $\Omega=[-40p, 40p]\times[-20 \sqrt{3}p, 20 \sqrt{3}p]$.}
	\label{fig:stress4}
\end{figure}

To understand the behavior observed in figure~\ref{fig:merging} and further characterize the effects described by the model, we focus on an idealized case, namely the growth of a small seed and the interaction with its periodic images in a domain hosting an integer number of the unit cell at $T_0$. The growing crystal is expected to interact with its periodic images. As shown in figure~\ref{fig:stress4}(a), with $\alpha^{\rm th}=0$, no merging occurs, and tensile stress is observed (see figure \ref{fig:stress4}(c)). The latter builds during the very first stages of the growth and decreases up to a given value when the temperature equilibrates to $T_0$. Note that no significant differences are observed in this regard with ($\gamma\neq 0$) and without ($\gamma=0$) the coupling with the heat flux (see figure~\ref{fig:stress4}(c)), so this effect is present in the classical PFC model. With $\alpha^{\rm th}=1.0$, we consider two cases corresponding to two different initialization of the average density. We consider $A_1=A$, for which $T<T_0$ in the crystal, and $A_2=1-A$, for which $T>T_0$ in the crystal, such to have opposite $T-T_0$ contributions (see also discussion of figure~\ref{fig:longTimeSim}). These two cases show different behaviors. The former case features lower tensile stress due to a negative contribution of the temperature-induced lattice expansion, partially compensating for the tensile stress observed during growth. As a result, merging occurs, and smaller residual stress is observed at later stages. The latter, instead, features an additional tensile-stress contribution due to the increase in the temperature, and no merging is observed. In this case, similar tensile stress is observed owing to the equilibration of the temperature at $T_0$ for an isolated crystal with a shape very similar to the $\alpha^{\rm th}=0$ case.

 \begin{figure}[t]
	\includegraphics{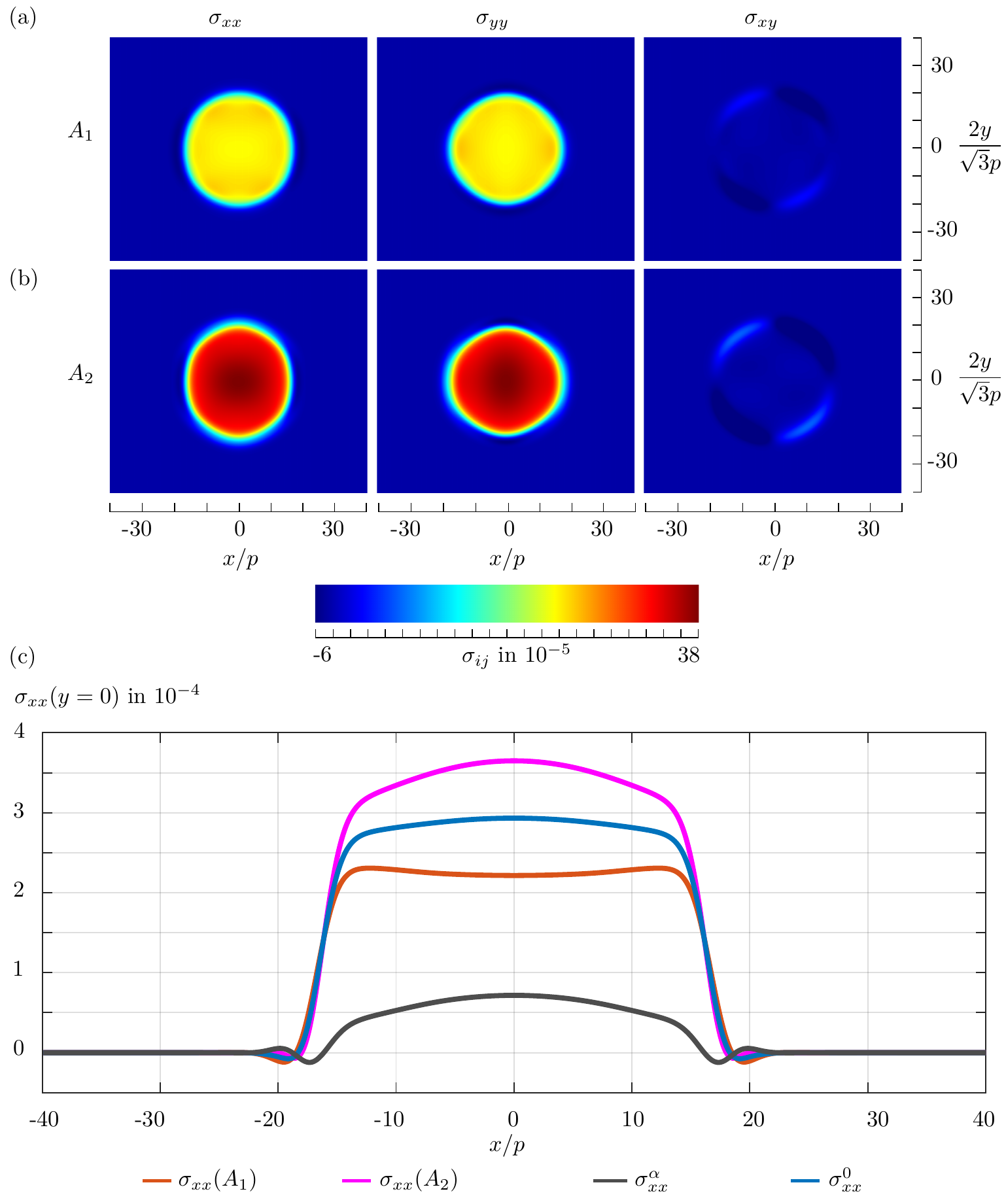}
	\caption{Stress field and thermal stress. A first stage of the simulations with $\alpha^{\rm th}=1.0$ in figure~\ref{fig:stress4} is considered. (a) $\sigma_{ij}$ for the case $A_1=0.849$. (b) $\sigma_{ij}$ for the case $A_2=0.151$. (c) $\sigma_{xx}(A_1)=(1/2)(\sigma_{xx}^0-\sigma_{xx}^\alpha)$ and  $\sigma_{xx}(A_2)=(1/2)(\sigma_{xx}^0+\sigma_{xx}^\alpha)$ as well as their split into $\sigma_{xx}^\alpha$ and $\sigma_{xx}^0$ are compared along the x-axis crossing the center of the crystal.}
	\label{fig:stress3}
\end{figure}

Exploiting the simulations initialized with both $A_1$ and $A_2$ we may provide an estimation for the thermal stress. The mechanical stress for $\alpha^\mathrm{th}=0$, namely $\sigma^0_{ij}$ may include different contributions in general, e.g., effects of an applied load and specific boundary conditions. Under similar conditions but with $\alpha^\mathrm{th}=1.0$ an additional contribution resulting from the dependence of the equilibrium lattice spacing on the temperature field, $\sigma^\alpha_{ij}$, is expected such that $\sigma_{ij}=\sigma_{ij}^0+ \sigma_{ij}^\alpha$. We point out that $\sigma_{ij}^0$ is not necessarily equivalent for $\alpha^\mathrm{th}=1.0$ and $\alpha^\mathrm{th}=0$ as the growth dynamics is different. However, it is expected to be similar for early stages of the growth where the morphology do not differ significantly, as we consider here. An estimation of $\sigma_{ij}^\alpha$ may be then given by defining
    $\sigma_{ij}(A_k)=\sigma_{ij}^0(A_k)+  \sigma_{ij}^\alpha(A_k)$, with $k=1,2$, and $\sigma_{ij}^\alpha(A_2)  \approx -\sigma_{ij}^\alpha(A_1) \eqqcolon \sigma_{ij}^\alpha$, reflecting the behavior of the temperature in the solid. We may then compute
$2\sigma_{ij}^\alpha = \sigma_{ij}(A_2) - \sigma_{ij}(A_1)$ assuming $\sigma_{ij}^\mathrm{0}(A_1)\approx\sigma_{ij}^\mathrm{0}(A_2)=(1/2)[\sigma_{ij}(A_1)+\sigma_{ij}(A_2)]$. In figure~\ref{fig:stress3} we show the mechanical stress components for the simulations initialized with average densities $A_1$ and $A_2$, and address their splitting into $\sigma_{ij}^0$ and $\sigma_{ij}^\alpha$ for an early stage of the simulations reported in figure~\ref{fig:stress4} with $\alpha^{\rm th}=1.0$.
Independently on the value of $\alpha^\mathrm{th}$, the stress vanishes in the liquid phase, while its values changes for $A_1$ and $A_2$, owing to the different temperatures, see Fig~\ref{fig:stress3}(a)--(b). More insights are given in Fig~\ref{fig:stress3}(c), reporting $\sigma_{xx}(A_1)$ and $\sigma_{xx}(A_2)$ along the x-axis crossing the center of the crystal as well as the computed $\sigma_{ij}^0$ and $\sigma_{ij}^\alpha$. $\sigma_{ij}^0$ closely resembles the value obtained for $\alpha^\mathrm{th}=0$ case. Importantly, opposite effects are obtained for $T<T_0$ and $T>T_0$ in the solid obtained with $A_1$ and $A_2$, respectively. A positive (negative) $\sigma_{ij}^\alpha$ is obtained for $T>T_0$ ($T<T_0$) in the crystalline phase, consistently with the expected thermal stress. A small region with opposite behavior at the solid-liquid interface may be ascribed to the features of the solidification process \cite{Galenko2011,Galenko2015} and includes the effects induced by changes in the temperature field. It is worth noting that considering a proper separation of the timescales for elastic relaxation and diffusive dynamics, eventually controllable through parameters, would account for a complete model allowing for the exploration of different relaxation regimes \cite{StefanovicPRL2006,HeinonenPRL2016,SkaugenPRL2018,skogvoll2021stress}. However, the purely diffusive dynamics considered here, delivering a slow elastic relaxation as encoded in the standard PFC, is instrumental in interpreting the different quantities entering the model, which may be extended without major adaptations of the aspects discussed above.

\section{Conclusion}
\label{sec:conclusions}
We introduced a PFC model of crystal growth accounting for heat fluxes and thermal expansion. As a starting point, we considered a thermodynamically consistent approach coupling the evolution of the density field with a diffusion-reaction equation for the temperature field. Then, a parameter controlling the periodicity of the density, dependent on a temperature field has been set consistently with the classical law of thermal expansions in solids. We assessed this description by comparison with a known analytic solution of a prototypical system, namely addressing the Eshelby inclusion problem in the PFC model. Moreover, we showed the model's capabilities by numerical simulations performed with a simple but efficient numerical scheme. An overview of how the parameters entering the model control the growing morphologies is provided, focusing on the dendritic growth regime. Importantly, we characterize thermal stress effects in the system and the change of spacing induced by temperature changes. This is found to affect the merging process of growing crystals.

This work sets the ground for a comprehensive approach to crystal growth at diffusive time scales, retaining microscopic features. Perspective extensions include the modeling and simulations of the growth of faceted crystals and dendrites in open systems beyond closed ones enforced here by periodic boundary conditions and accounting for the latent heat of solidification. Like classical PFC and phase-field models, the approach presented here could be readily applied to the study of three-dimensional systems with the aid of efficient numerical methods \cite{TANG2011146,Yamanaka2017}. However, for this purpose, the so-called amplitude expansion of the PFC model \cite{Goldenfeld2005,salvalaglio2022coarse} may also represent an ideal framework as it features an additional spatial coarse-graining and may enable large-scale simulations. Additionally, considering an explicit modeling of elastic relaxation may allow for a better description of competitive time scales, in particular concerning elastic relaxation and diffusive dynamics \cite{StefanovicPRL2006,HeinonenPRL2016,SkaugenPRL2018,skogvoll2021stress}, also in the presence of heat flux.

\section*{Acknowledgements}
MP and MS acknowledge support by the German Research Foundation (DFG) under Grant No.~SA4032/2-1. AV acknowledges support by the German Research Foundation (DFG) within SPP1959 under Grant No.~Vo899/20-2. SMW gratefully acknowledges support from the US National Science Foundation under grant NSF-DMS 2012634. The authors acknowledge useful discussion with Ken R. Elder. Computing resources have been provided by the Center for Information Services and High-Performance Computing (ZIH) at TU Dresden, and by J\"ulich Supercomputing Center under grant PFAMDIS.

\section*{References}
\bibliographystyle{iopart-num-mod} 
\bibliography{references.bib}

\end{document}